\documentclass[letterpaper,conference]{IEEEtran}
\IEEEoverridecommandlockouts
\usepackage{cite}
\usepackage{graphicx}
\usepackage{textcomp}
\usepackage[table]{xcolor}
\def\BibTeX{{\rm B\kern-.05em{\sc i\kern-.025em b}\kern-.08em
    T\kern-.1667em\lower.7ex\hbox{E}\kern-.125emX}}
    
\usepackage{xspace}
\usepackage{multirow}
\usepackage{graphicx}
\usepackage{adjustbox}
\usepackage{tabularx}

\definecolor{c0}{HTML}{1F77B4} 
\definecolor{c1}{HTML}{FF7F0E} 
\definecolor{c2}{HTML}{2CA02C} 
\definecolor{c3}{HTML}{D62728} 
\definecolor{c4}{HTML}{9467BD} 
\definecolor{c5}{HTML}{8C564B} 
\definecolor{c6}{HTML}{E377C2} 
\definecolor{c7}{HTML}{7F7F7F} 
\definecolor{c8}{HTML}{BCBD22} 
\definecolor{c9}{HTML}{17BECF} 

\rowcolors{2}{white}{c7!25}
\usepackage{booktabs}

\usepackage{listings}
\lstdefinelanguage{jobConf}
{
  keywords={
    first,
    FLOW,
    CLIENT,
    REQS,
    TASK,
    ADDR,
    NAME,
    VALUE,
    ID,
    STR,
    PARAM,
    last
  }
  sensitive=true,
  basicstyle=\bfseries\ttfamily\small,
  captionpos=b,
  keepspaces=true,
  numbers=none,
  morestring=[b]",
  keywordstyle=\color{c0},
  stringstyle=\color{c2},
  showspaces=false,
  showtabs=false,
  breaklines=true,
  showstringspaces=false,
  breakatwhitespace=true,
}

\lstdefinelanguage{job}
{
  basicstyle=\bfseries\ttfamily\small,
  morecomment=[s][\color{c0}]{<}{>}, 
  captionpos=b,
  keepspaces=true,
  numbers=none,
  showspaces=false,
  showtabs=false,
  breaklines=true,
  showstringspaces=false,
  breakatwhitespace=true,
}

\newcommand{\dtn}{OPP\-LOAD\xspace}

\begin{document}


\setlength{\textfloatsep}{0.8\baselineskip}

\title{\dtn: Offloading Computational Workflows in Opportunistic Networks}

\author{
	\IEEEauthorblockN{
		Artur Sterz\IEEEauthorrefmark{1}, 
		Lars Baumg\"{a}rtner\IEEEauthorrefmark{2}, 
		Jonas H\"{o}chst\IEEEauthorrefmark{1}\IEEEauthorrefmark{3}, 
		Patrick Lampe\IEEEauthorrefmark{1}\IEEEauthorrefmark{3},
		Bernd Freisleben\IEEEauthorrefmark{1}\IEEEauthorrefmark{3}
	} \\
	\IEEEauthorblockA{
		\IEEEauthorrefmark{1}\textit{Dept. of Electrical Engineering \& Information Technology, TU Darmstadt, Germany}\\
		E-mail: \{artur.sterz, jonas.hoechst, patrick.lampe, bernd.freisleben\}@maki.tu-darmstadt.de
	}
	\IEEEauthorblockA{
		\IEEEauthorrefmark{2}\textit{Dept. of Computer Science, TU Darmstadt, Germany}
		E-mail: baumgaertner@cs.tu-darmstadt.de
	}
	\IEEEauthorblockA{
		\IEEEauthorrefmark{3}\textit{Dept. of Mathematics \& Computer Science, Philipps-Universit\"{a}t Marburg, Germany} \\
		E-mail: \{lbaumgaertner, hoechst, lampe, freisleb\}@informatik.uni-marburg.de
	}
}

\maketitle

\begin{abstract}
Computation offloading is often used in mobile cloud, edge, and/or fog computing to cope with resource limitations of mobile devices in terms of computational power, storage, and energy.
Computation offloading is particularly challenging in situations where network connectivity is intermittent or error-prone.
In this paper, we present \dtn, a novel framework for offloading computational workflows in opportunistic networks.
The individual tasks forming a workflow can be assigned to particular remote execution platforms (\emph{workers}) either preselected ahead of time or decided just in time where a matching worker will automatically be assigned for the next task.
Tasks are only assigned to capable workers that announce their capabilities.
Furthermore, tasks of a workflow can be executed on multiple workers that are automatically selected to balance the load.
Our Python implementation of \dtn is publicly available as open source software.
The results of our experimental evaluation demonstrate the feasibility of our approach.
\end{abstract}

\begin{IEEEkeywords}
Offloading, Opportunistic Networks, Workflows 
\end{IEEEkeywords}

\section{Introduction}
\label{sec:introduction}

Opportunistic networking is useful for communication in scenarios where no infrastructure is available, if network connectivity is intermittent or error-prone.
This is achieved using a store, carry and forward approach to transmit
bundles hop-to-hop, from source to destination.
Opportunistic networking can help first responders and victims in disasters,
inhabitants in rural areas, and researchers in environmental monitoring of natural habitats to exchange data without relying on a working communications infrastructure \cite{gardner2011serval, baumgartner2018environmental,conti2010opportunistic}.
Since mobile devices used in such scenarios typically have
limited computational power, storage, or energy, offloading computational
tasks can reduce load on initiating devices or even enable task execution, e.g., if specialized hardware is required~\cite{kumar2013survey}.
Face detection in disaster scenarios could help first responders to save resources for essential communication~\cite{conti2010opportunistic}, and environmental monitoring with mobile sensor nodes is a current research topic~\cite{baumgartner2018environmental}.

\begin{figure}[t]
    \centering
    \includegraphics[width=.70\columnwidth]{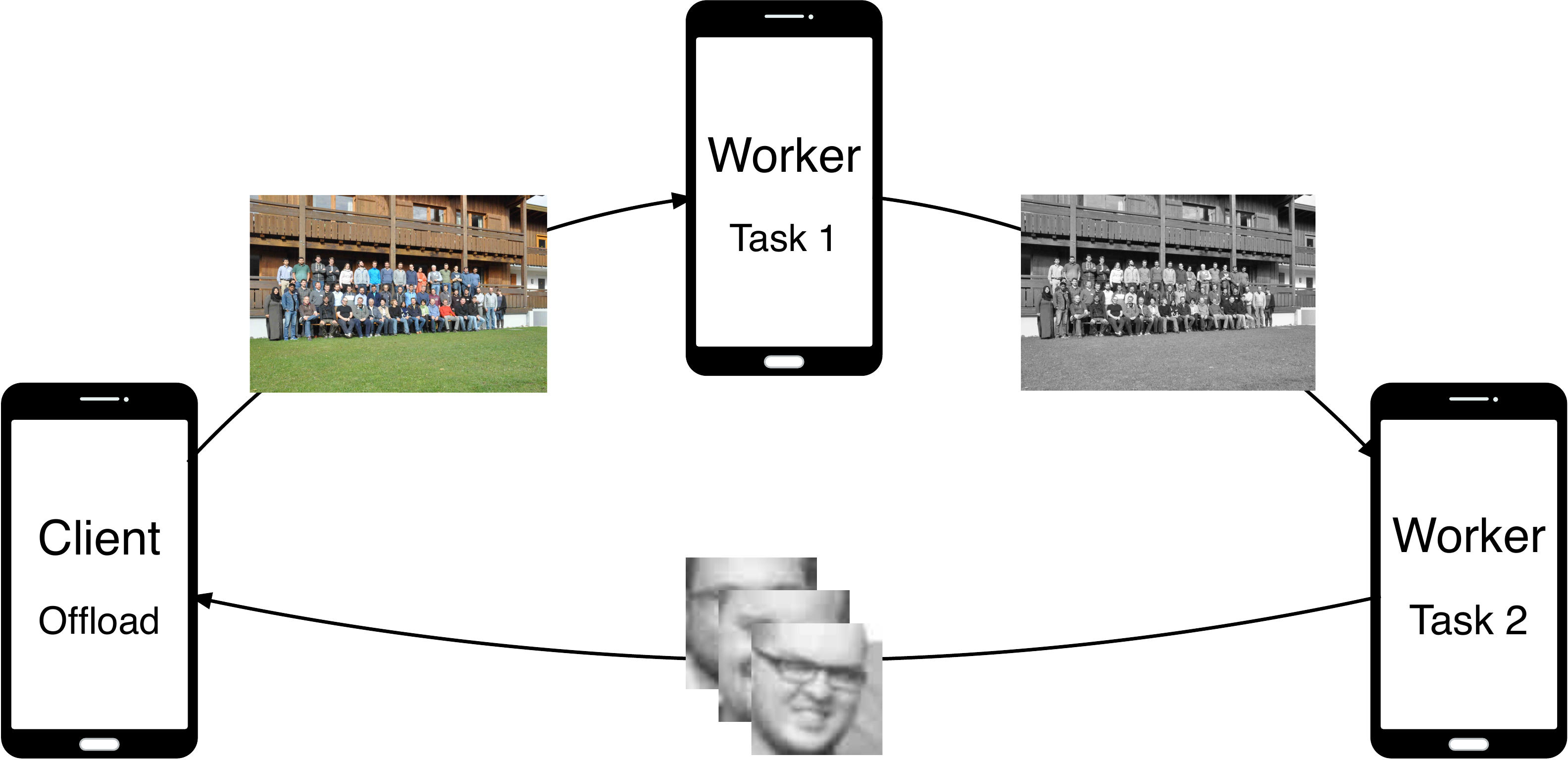}
    \caption{Illustrative example: executing a workflow on two workers.}
    \label{fig:motivation}
\end{figure}

In this paper, we present \dtn, a novel framework for offloading computational workflows in opportunistic networks.
Fig.~\ref{fig:motivation} shows an example for offloading a simple face detection workflow where Task 1 converts an image to grayscale, and Task 2 extracts faces and returns them to the client~\cite{lampe2017smartface}.
Using \dtn, \emph{clients} can assign the individual tasks of a workflow to particular remote execution platforms (\emph{workers}) ahead of time or can leave the assignment open, i.e., each worker will search for another suitable worker just in time for the next task using our novel worker assignment algorithm, while both modes can be mixed in a workflow.
Workers publish their capabilities and resources (available memory, remaining battery capacity, etc.), i.e., clients will only select capable workers.
Furthermore, workflow tasks can be executed on multiple workers that are automatically selected to balance the overall load, based on a folded standard normal distribution and an innovative worker ranking system, where workers are rated based on their available resources, their capabilities, and the proximity to the calling client/worker, resulting in a novel approach ensuring that load on workers is distributed fairly in the network. No permanent connection to the worker nodes is needed. Depending on node mobility, results can still be delivered even after longer periods of isolation due to a disruption-tolerant networking (DTN) underlay.

To show the feasibility of our approach, we emulated up to 30 highly mobile nodes in different experimental settings, showing that the success rate of offloading increases by up to 40\% with negligible overhead.
Our Python implementation\footnote{https://github.com/umr-ds/OPPLOAD} and all artifacts\footnote{https://github.com/umr-ds/OPPLOAD-experiments}\footnote{https://ds.mathematik.uni-marburg.de/oppload/oppload\_results.tar.gz} of this paper are publicly available.

The paper is organized as follows. Section~\ref{sec:related_work} discusses related work. Section~\ref{sec:design} presents the design of \dtn, Section~\ref{sec:implementation} covers its implementation. Section~\ref{sec:evaluation} presents an experimental evaluation of \dtn. Section~\ref{sec:conclusion} concludes the paper and outlines areas for future work.

\section{Related Work}
\label{sec:related_work}

\subsection{Workflow-based Approaches}

To offload tasks to other mobile devices, Serendipity splits each task into smaller tasks that are either offloaded or not if no worker is found~\cite{shi2012serendipity}.
In a mobile cloud computing scenario, Ahn et al. \cite{ahn2018competitive} start the execution of tasks locally and offload them to suitable cloudlets.
Ravi and Peddoju ~\cite{ravi2018mobile} present an offloading algorithm  where an application is partitioned into clusters containing tasks to decide whether to offload, based on a density-based clustering algorithm.

Although these proposals follow a workflow-based approach, they do not have a worker selection algorithm to distribute the load fairly in the network and/or they are not suitable for opportunistic networks.

\subsection{Proximity-based, Movement-based, and Social Approaches}
COMET~\cite{gordon2012comet} is a framework for offloading parts of applications to neighboring nodes to speed up their execution.
Mtibaa et al.~\cite{mtibaa2013towards} propose a framework where a task is offloaded to mobile devices that belong to the same social context, e.g., the same household or a group of first responders in a disaster scenario.

Wang et al.~\cite{wang2014mobility} present an offloading scheme for opportunistic networks where mobility patterns are analyzed to estimate the number and duration of contacts for the offloading decision.
Zhang et al.~\cite{zhang2015offloading} consider the load of a device, the availability of cloudlets, and user mobility to maximize the probability of successfully offloading tasks.
Honeybee~\cite{fernando2016computing} includes a work sharing algorithm that employs nearby nodes to execute tasks based on job stealing.

These publications either focus on a single aspect (e.g., movement/proximity of nodes or social relationships), or they are designed for cloudlet scenarios and thus are not suitable for opportunistic networks.
Additionally, most of these approaches do not follow a workflow-based approach and offload only entire tasks, without splitting them into smaller tasks.

\subsection{Offloading in Cloud Environments}
Deng et al.~\cite{deng2015computation} decide for each task of a workflow whether it should be offloaded to the cloud or executed locally, based on the capabilities and the movement of nodes.
Chatzopoulos~\cite{chatzopoulos2016have} use an incentive mechanism where users have to define how many resources they are willing to spend for executing offloaded tasks.
Chowdhury et al.\cite{chowdhury2018context} migrate tasks between cloud, mobile devices, or robots by considering energy, latency, and task execution deadlines.

All these works are designed for cloud environments and are therefore not optimized for resource savings in opportunistic networks with mobile devices.

\subsection{Mobile Cloud, Edge, and Fog Environments}
Fan et al.~\cite{fan2018computation} present an approach where a base station in a mobile cloud scenario can either execute an offloaded task itself or further offload it to another base station.
Using a fuzzy decision engine, Flores et al.~\cite{flores2013adaptive} consider multiple criteria like CPU power to decide whether a task should be offloaded to a mobile cloud server.
Yang et al.~\cite{yang2013framework} offload computations in mobile cloud scenarios to maximize the throughput of applications.
Chen et al.~\cite{chen2016efficient} formulate a game theoretic approach for offloading tasks in a mobile cloud scenario.
Bellavista et al.~\cite{bellavista2017migration} present a computation offloading approach, where tasks are offloaded to mobile edge cloud instances and the results are return over the same node or a different one, if the user has moved in the meantime.
Zhang et al.~\cite{zhang2018real} introduce a task allocation scheme where social sensing applications are offloaded to edge servers to maximize a node's payoff by saving energy.
Yang et al.~\cite{yang2018network} propose an algorithm to offload tasks to a nearby edge server.

These approaches assume the availability of a mobile cloud, cloudlets, or similar technologies.
In addition, neither worker capabilities, nor highly unreliable networks, nor workflow-based execution to preserve resources are taken into account.

\subsection{Other Approaches}
Funai et al.~\cite{funai2016mobile} present an approach that minimizes energy consumption by offloading computations across multiple hops in an ad-hoc network.
Zanni et al.~\cite{zanni2017automated} propose an approach to split arbitrary Android apps into smaller tasks that can be offloaded.
Sterz et al.~\cite{sterz2017dtn} present a framework for remote procedure calls in disruption-tolerant networks with separated control and data channels to cope with short contact durations.
Internet-of-Things devices use more capable devices that are reachable within one hop to execute a task~\cite{elazhary2018w}.
Feng et al.~\cite{feng2018computation} present an approach where mobile devices offload tasks to other mobile devices via cellular base stations without prior knowledge of the devices' resources.

These approaches are either not suitable for opportunistic networks and faulty situations, or they only consider a very limited scope of capabilities and worker selection.
Furthermore, most of them do not handle workflows but single tasks only, which is not suitable for scenarios where mobile devices are the main execution platforms.

Finally, to the best of our knowledge, there is no previous work that takes all these parameters into account, introduces a transparent workflow-based computational task offloading algorithm for multi-hop opportunistic networks, and provides an open source proof-of-concept implementation.

\section{\dtn's Design}
\label{sec:design}

\subsection{Workflow-based Computations}
\label{sub:workflow-based-computations}

\dtn supports \emph{workflow-based computations} where a client defines a workflow that consists of a chain of tasks.
The client assigns each task to a worker, and \dtn will take care of the execution order, even in unpredictable network situations.
Furthermore, \dtn transparently passes inputs and outputs between the different tasks of a workflow.
Connectivity is achieved using protocols for disruption-tolerant networking (DTN), while we assume that the communication overhead in terms of CPU and memory resources for remote execution is negligible~\cite{baumgartner2016experimental}.

\subsection{Worker Addressing}
\label{sub:worker-addressing}
\dtn supports two worker addressing modes: \emph{Ahead of Time (AoT)} and \emph{Just in Time (JiT)}.
This makes it possible to select the best suitable and available worker for each task, based on the user's preferences and the network environment.

\subsubsection{Ahead of Time}
Using AoT addressing, the client assigns a task to a worker explicitly.
It is possible to select a different worker for each task, as well as the same worker for different tasks.
This mode exists mainly for two reasons.
Privacy-sensitive tasks should be executed on known and trusted workers.
Furthermore, worker operators might give certain guarantees, e.g., to stay in the network or to always execute a task, even under heavy load.

\subsubsection{Just in Time}
In JiT mode, workers publish all services they offer periodically by broadcasting them into the network.
These offers are stored on every node locally, where workers are searched from.
Since in opportunistic networks nodes can appear and disappear frequently from the network, these offers are only valid for a certain time period, depending on the dynamics of the network.

\begin{figure}[t]
    \centering
    \includegraphics[width=.90\columnwidth]{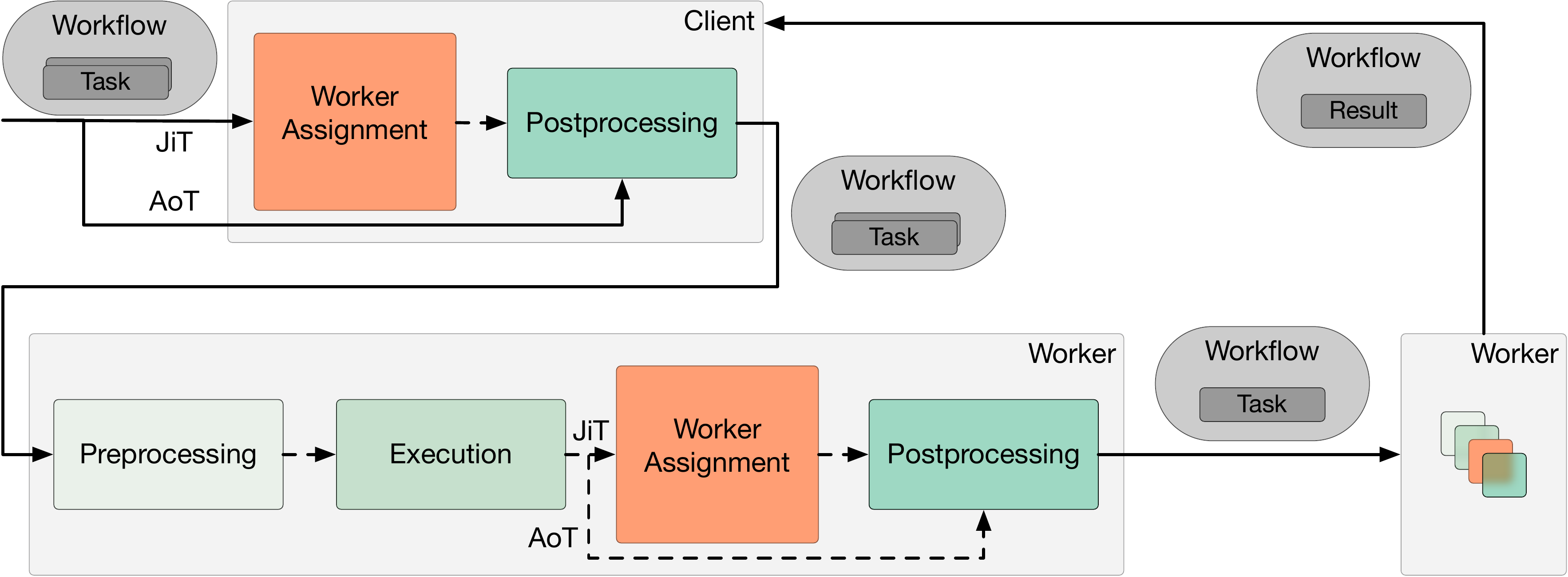}
    \caption{Architecture of \dtn client and worker showing a possible workflow with Ahead of Time (AoT) or Just in Time (JiT) worker assignment.}
    \label{fig:design}
\end{figure}

If a client does not assign a task to a specific worker, \dtn will transparently chose a suitable worker by passing the workflow description through a number of steps that are part of worker assignment, as shown in Fig.~\ref{fig:design}.
During the first step of worker assignment, a worker with an offer for executing a desired task will be searched in the local database.
If the search is successful, the task will be executed on this worker.
This mode is helpful when it is not clear whether a worker is available for a task.

\subsection{Worker Capabilities}
\label{sub:worker-capabilities}

Workers announce their \emph{capabilities} and \emph{available resources}, such as CPU load, available memory, and other metrics.
Additionally, workers announce available special hardware or other properties that help executing specific tasks better than other workers, e.g., face detection in images is more energy efficient on a GPU that may not be available on all workers.
The time interval for periodic capability announcements matches the dynamics of the network.
The more dynamic a network is, the more often the capabilities are broadcast.
In the second step of worker assignment using JiT addressing, these capabilities are taken into account.
Task requirements specified by the client are compared to the capabilities published by the workers to select capable workers.

\subsection{Worker Assignment}
\label{sub:worker-selection}
During worker assignment, multiple capable workers may be available.
Therefore, we have developed a novel \emph{worker assignment algorithm} that distributes the workload fairly in the network on multiple workers and selects nearby and powerful workers.
Instead of simply selecting a random worker or the worker with the most recent offer, we introduce a worker rating scheme based on different weighted metrics.
The user has to estimate, e.g., CPU, memory, or disk space requirements for a task.
Additionally, the rating scheme also keeps the tasks spatially close to the calling client.
Therefore, the geographical distance between the two involved nodes is a metric of the rating.
During worker assignment, the client will calculate for each capable worker how well it satisfies each requirement of a task by dividing the capabilities published by workers by the given requirements for every metric.
By applying this rating scheme, the best capable worker based on the local knowledge is selected.
However, this can lead to an unfair load distribution in the network, where nearby and powerful workers could be disadvantaged, since they would always be chosen.
Therefore, a worker is selected from the sorted list of workers based on their rating following the folded standard normal distribution.
This ensures that a nearby and powerful worker will be selected with a high probability, but the load is also distributed to different workers, leading to a fair workload distribution approach.

\subsection{Error Handling}
\label{sub:dealing-with-errors}

Bundle delivery in opportunistic networks cannot be guaranteed.
If a worker disappears in \dtn before it could execute an assigned task, the client would wait infinitely long.
Therefore, users can specify a time-to-live (TTL) for a workflow.
This has two implications.
First, the client stops waiting for the results after the TTL has expired, making it possible to re-issue the workflow.
Second, a worker will not execute a task if the TTL is expired, which preserves resources on workers.
This ensures a defined behavior in cases where no result can be retrieved in time.

If errors occur in conventional networks, clients can be notified immediately to handle the error appropriately.
In opportunistic networks, this is not necessarily possible due to potentially poor network conditions.
Thus, \dtn handles three classes of error.
The first error class is a \emph{task execution error}.
These errors occur during the execution of the task itself.
The offloaded task can implement error and exception handling on its own and provide error messages and stack traces, which \dtn will deliver to the client.
The second error class is a \emph{worker selection error}.
These errors occur if the execution of a task was successful, but a worker cannot find a subsequent worker during the assignment.
The third error class is a \emph{worker calling error}.
These errors can occur in different situations, such as when the worker is no longer capable to execute the task or if it is not offering the service and was called by mistake in AoT mode.
Error handling for these errors depends on the addressing mode.
If the worker on which the error occurred was selected in JiT mode, it will inform the prior worker about the error, which will retry to assign the task to a capable worker one more time.
If the second try also fails or the worker was chosen in AoT mode, the client will be informed about the error using the same communication mechanisms as before.
The client is then responsible to handle the error appropriately.
After an offloaded task finishes or an error occurs, \dtn will clean up all involved files and bundles across all workers to save storage.

\section{Implementation}
\label{sec:implementation}

We implemented \dtn based on the bundle store implementation, \emph{Rhizome}, of the Serval Mesh~\cite{gardner2011serval}, which uses a simple epidemic DTN routing protocol.
\dtn is written in Python and uses Rhizome's RESTful API for handling all network-related duties.
In previous work, we have conducted an in-depth evaluation of Serval in various experiments with different network setups and usage patterns~\cite{baumgartner2016experimental}.

\subsection{Offering a Service}
\label{sub:offering-a-service}

Workers offer a service by a name, an arbitrary number of parameters, and an executable that should be executed on the worker.
Any executable that runs on the underlying operating system can be used, e.g., Python programs, or compiled binaries.
Every worker periodically publishes the definitions of its services, and clients will then use these offers for the JiT worker assignment.
In addition to the service offers, workers also announce their capabilities as key-value pairs that are published together with the service offers to reduce the network overhead.

\subsection{Executing a Workflow}
\label{offloading-a-task}

To execute a workflow, a user splits it into tasks to be executed across multiple workers.
All tasks have to be described in a workflow description containing the desired worker (either AoT or JiT), the name of the task, and all required parameters, for each task.
A workflow description must include at least one task.
This workflow description has to be provided to the \dtn client that handles the remaining parts transparently.

A workflow description has the following form.
First, a task has to be assigned to a worker, which can be an address for AoT mode or a placeholder indicating that JiT mode should be used.
Then, the name of the service to be executed has to be given, followed by all parameters.
Using another placeholder indicates that the output of a task should be the input for the next task.
Each task can only have one result, and the placeholder is allowed only once per task.
Finally, a task can have requirements that are only used during assignment for this particular task.
After specifying the workflow, \dtn will assign a worker to the first task, if applicable.
The first step is to rank all workers, which is based on the requirements, as introduced in Section~\ref{sub:worker-selection}.
For each metric, a weighted rank is calculated and summed up, using the weight and the requirements as well as the worker's capability for the particular metric.
Workers are sorted based on their ranking, and a random worker is selected based on the folded normal distribution with location parameter $\mu = 0$ and scale parameter $\sigma = 1$.
All files required for a task, the workflow description itself, and task results or errors will be packed into an archive that will be sent as an encrypted bundle to the selected worker.
By packing everything in a single archive, fragmentation in transmission is avoided, and a worker is guaranteed to have everything required for processing the task.

When the offloaded task arrives, the worker starts preprocessing by unpacking the archive and parsing the workflow description.
It will check whether it is capable of executing the assigned task, since the capabilities could have changed during the transmission due to network delays.
If the worker is capable, the service will be executed.
After the service finishes, the worker will replace the parameter placeholder of the next task in the description with the result of its execution.
Finally, the worker assigns a next worker if required, packs everything into an archive, and passes it on.

When the final task is executed, the last worker will return the result to the client that will then trigger a network cleanup.
This is achieved by having the workers remove their payloads, and it is finished when the final result is removed.
If an error occurs, the worker will stop further execution, pack all intermediate files including the error log into an archive and return them as an error bundle to the client, which will raise an exception.
The client is responsible to handle the exception appropriately, e.g., re-execute the workflow.

\section{Evaluation}
\label{sec:evaluation}

\begin{figure*}[t]
    \centering
    \adjincludegraphics[width=\textwidth, Clip={0.0\width} {0.0\height} {0.0\width} {0.0\height}]{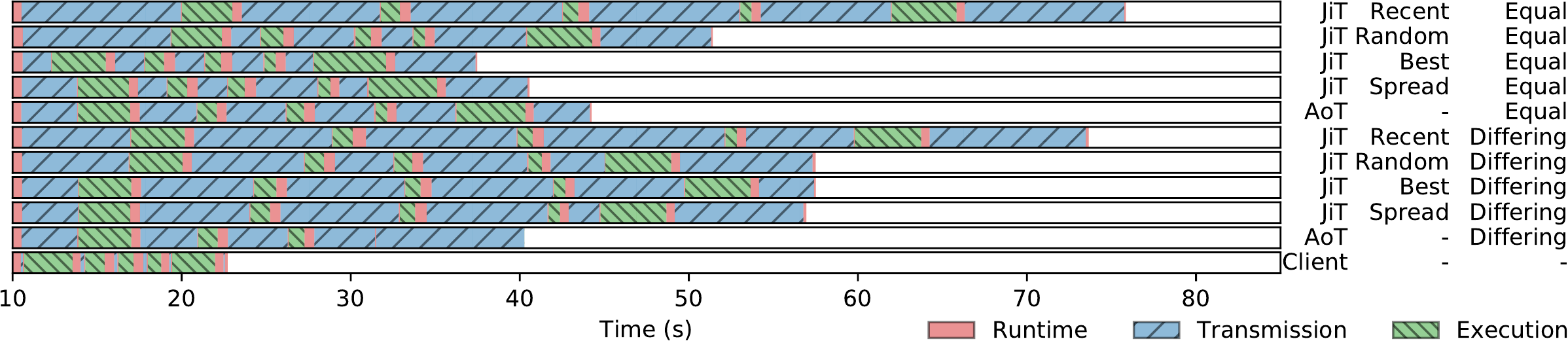}
    \caption{Exemplary overall workflow time in different configurations.}
    \label{fig:profiling-ring}
\end{figure*}

\subsection{Test Setup}
\label{sub:test-setup}

To evaluate \dtn in a realistic manner, the network emulation framework \emph{Common Open Research Emulator} (CORE) was used.
In contrast to simulation approaches like NS-3, CORE uses Linux namespaces to execute binaries and scripts natively, which gives us the opportunity to evaluate software and frameworks as close to reality as possible by still being able to scale the experiments easily~\cite{ahrenholz2008core}.

\subsubsection{Test Cases}
To evaluate \dtn, the algorithm of Lampe et al.~\cite{lampe2017smartface} for detecting faces in images on smartphones was adapted.
The workflow of this algorithm has five tasks.
The first task is to denoise an image.
The second task is to scale the image up by 10\% to increase the probability of fitting a possible face into the detection window.
The third task is to crop the image by 10\% to decrease the image size, which speeds up the detection time while maintaining a high detection accuracy.
The fourth task converts the colored image into an 8-bit grayscale image, which additionally speeds up face detection while maintaining the same detection accuracy.
The fifth task detects faces on the preprocessed image.
These five tasks are executed on five different workers in the network.
In every experiment, the bandwidth of the network links was set to 54 Mbit/s, and a delay of 20 ms was used.
All nodes were configured as workers.
In JiT mode, we compared four worker assignment algorithms: (i: \emph{recent})  selecting the worker whose offer arrived most recently, (ii: \emph{random}) selecting a worker randomly, (iii: \emph{best}) selecting the best available worker based on our rating, and (iv: \emph{spread}) using the algorithm  described in Section~\ref{sub:worker-selection}  to spread the load among the best available workers.
Since \dtn is designed for networks with mobile devices as workers, the weights for the worker rating were set to keep the tasks on nodes with high energy reserves and spatially close to the client.
Therefore, available energy and distance were weighted with 30\%, CPU with 20\%, and available memory and free disk space with 10\%.
We modelled energy using a virtual energy unit $e$.
These were used to model energy consumption for each task related to the task's execution time, meaning that the longer a task takes on average, the more $e$ is consumed.
In our experiments, a service offer from a worker was set to expire after 120 seconds, as described in Section~\ref{sub:worker-addressing}.
Finally, workers announced their capabilities every 2 seconds, since this is the sweet spot announcement interval, as shown by Baumg\"{a}rtner et al.~\cite{baumgartner2017speak}.

\subsection{Baseline Evaluation}
\label{sub:baseline-eval}
The baseline tests show how \dtn performs under good network conditions.
For these tests, twelve static nodes were arranged in a ring, where each node had exactly two neighbors
and only the first node was a client.
In AoT mode, workers were selected at the start of an experiment in the same order as they appear in the network, always skipping one node.
The same workers were used for all AoT experiments for comparability.
To evaluate the effect of worker capabilities, this setup was first executed with all workers equally capable of executing a task and a second time where we used the following capability distribution: 20\% (2) of the workers were capable with no constraints, 40\% (5) were also capable, but had less energy reserves, 30\% (3) could execute the task, but with limited capabilities (like little available memory) and 10\% (2) were not capable to execute the task at all.
The capabilities were modeled using available disk space, memory, CPU resources, and energy $e$, which was reduced according to the above description.
Since worker assignment requires randomness, our random number generator was initialized with 25 different seeds.
Finally, we executed the experiments also on the client to have a benchmark for comparison.

\subsubsection{Workflow Profiling}
To analyze the overhead of \dtn, workflow processing was split into three phases: (i) runtime (red) of the \dtn implementation, i.e., pre- and postprocessing and worker assignment in JiT tests, (ii) transmission time (blue) for transmitting the bundle, and (iii) execution time (green) of the task itself.
The colors refer to Fig.~\ref{fig:profiling-ring}.
The x-axis shows the workflow execution time, each bar denotes a specific configuration.

As shown in Fig.~\ref{fig:profiling-ring}, \dtn does not introduce significant processing overhead.
The workflows are offloaded from the clients 10~seconds after the start of the experiment.
Regardless of the test configuration, postprocessing and worker assignment require about 1 second, while preprocessing can be neglected.
The execution time depends on the task.
While scaling, cropping, and grayscaling only require about 2 seconds, denoising and detecting faces can take up to 6 seconds.

If AoT addressing is used in known topologies, users can estimate a workflow time range in which it finishes.
The downside is that if a worker is not capable of executing a task, the entire workflow will be stopped, as indicated by the second last bar in Fig.~\ref{fig:profiling-ring}.
Therefore, tasks should only be explicitly assigned in cases where no other option is desirable, or if a task must be handled by a specific worker.

The major overhead is introduced by transmitting the bundles across the network.
The last bar of Fig.~\ref{fig:profiling-ring} shows the same workflow executed on the client, thus no networking is needed.
The entire workflow needs about the same time as two to three tasks in the JiT tests, depending on the worker assignment.
Although overhead is introduced by network related operations, it can still be better to offload workflows than executing them locally.
First, the client may not be able to execute the tasks due to resource constraints or other limitations.
Second, the longer the tasks take to be executed, the more negligible the communication overhead becomes.
Finally, the decision whether to offload or not also depends on the number of hops between the offloading node and the worker, as indicated by Graubner et al.~\cite{graubner2018opportunistic}.
For \dtn, we assume that the user decides whether to offload during the creation of the workflow.

\begin{table}[t]
    \centering
    \caption{Average runtimes of workflow parts in the ring scenario in client-only tests and using AoT addressing.}
    \label{tab:runtimes_aot_client}
    \begin{tabularx}{\columnwidth}{Xrrrr}
    \hline
    Addr.        & Exec. (s)   & Runt.   (s) & Transm. (s)  & Total (s)     \\
    \hline
    Client       & 8.10 (0.21) & 3.55 (0.11) & 0.87 (-)     & 12.52 (0.32)  \\
    AoT          & 9.94 (0.26) & 3.77 (0.08) & 20.44 (0.13) & 34.15 (0.47)  \\
    \hline
    \end{tabularx}
\end{table}

\begin{table}[t]
    \centering
    \caption{Average runtimes of workflow parts in the ring scenario using JiT addressing and all four assignments.}
    \label{tab:runtimes_jit}
    \begin{tabularx}{\columnwidth}{Xrrrr}
    \hline
    Assign. & Exec. (s)    & Runt. (s)   & Transm. (s)   & Total (s)     \\
    \hline
    Recent  & 9.65 (0.26)  & 3.89 (0.09) & 50.94 (10.20) & 64.48 (10.50) \\
    Random  & 9.82 (0.16)  & 3.93 (0.09) & 32.60 (4.27)  & 46.35 (4.25)  \\
    Best    & 10.02 (0.28) & 3.94 (0.08) & 23.54 (9.63)  & 37.49 (9.99)  \\
    Spread  & 9.95 (0.20)  & 3.95 (0.09) & 24.05 (6.82)  & 37.94 (7.11)  \\
    \hline
    \end{tabularx}
\end{table}

Tables~\ref{tab:runtimes_aot_client} and~\ref{tab:runtimes_jit} show the average time needed for the parts of a workflow (the numbers in brackets show the standard deviation) in seconds.
Table~\ref{tab:runtimes_jit} indicates that the overall workflow time highly depends on the worker assignment in the JiT experiments.
The recent worker assignment with an average of about 64.48 seconds requires the longest time, due to the long distance between the nodes, since their offers take longer to reach the client and thus arrive more recently.
The standard deviation is also relatively high with more than 10 seconds, indicating long running tasks and differing results.
The random worker assignment achieves better results with about 46.35 seconds on average and a deviation of 4.25 seconds, since closer workers are chosen.
Always selecting the best available worker leads to significantly lower workflow times, requiring about 37.49 seconds, but with a standard deviation of 9.99 seconds.
Finally, using the spread assignment algorithm, the workflow time does not significantly differ from the previous assignment algorithm, using about 37.94 seconds, but has a better standard deviation of 7.11 seconds.
If all workers are equally capable, the workflow times using the best worker or the spread algorithm do not differ.
This shows clearly that in terms of workflow time, the algorithm using the best workers and our spread approach outperform the other approaches.
But since not all workers are equally capable in the different capability tests, tests using the best worker have a broader standard deviation, since the capable workers are further away in the topology.
This means that always using the best worker is slightly faster than using the spread algorithm, but is more unpredictable in how long the execution of a workflow will take, since the very best workers will be worn up and worse workers have to be chosen consequently.
Therefore, we propose our spread algorithm as the best available solution.
Executing a workflow locally at the client would only require execution time and runtime, since the networking part is not needed.
As shown in Table~\ref{tab:runtimes_aot_client}, the total execution time is about 12.52 seconds and is pretty stable with only about 300 ms deviation.
Finally, the AoT mode needs about 34.15 seconds in total and is also stable with only about 400 ms deviation.
Since in AoT mode a worker is always two hops away from the next hop,
the transmission is even faster than using JiT mode with the best worker assignment.

\subsubsection{Worker Load Distribution}

\begin{figure}[t]
    \centering
    \includegraphics[width=.90\columnwidth]{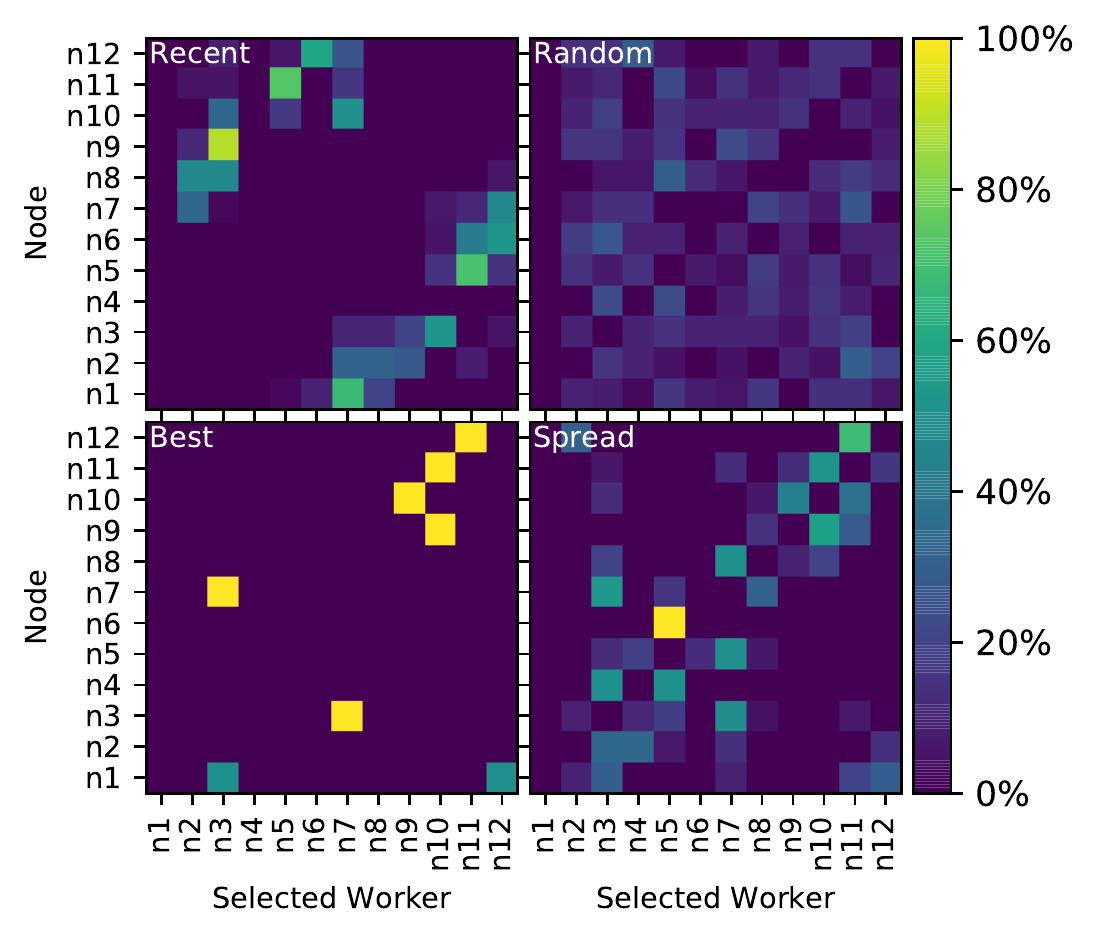}
    \caption{Selected workers in ring JiT scenarios.}
    \label{fig:heatmap-ring}
\end{figure}

Fig.~\ref{fig:heatmap-ring} shows the worker load distribution in all four worker assignment algorithms using JiT mode.
On the y-axis, the calling nodes are shown, whereas on the x-axis the assigned worker is denoted.
The lighter the color, the more often a particular client selected a particular worker.

The recent selection approach spreads the load over particular nodes, but almost always selects a worker on the opposite side of the network, leading to long-running workflows.
Using a random worker, the workload is distributed on nearly all available workers.
Although this leads to a fair load distribution, the profiling analysis shows that this approach does not necessarily give the fastest workflow execution times.
Additionally, tasks are sent to spatially far away workers, leading to the same problems of long transmission times and network splits in mobile networks as in the recent approach.
Always using the best available worker keeps the workflow execution spatially close, and the overall runtimes are the lowest achievable, but with an unfair load distribution, which disadvantages close and powerful workers over others that are also able to execute a task.
In dense networks with a high offloading frequency, this could lead to overloaded nodes and empty batteries, which in the end would be less beneficial for the overall performance.
Finally, our approach spreading the load on the best workers leads to the best overall results.
Close and powerful workers are preferred over others, while less powerful workers also have chances to be selected.
Overall, as previously shown in Table~\ref{tab:runtimes_jit}, the workflow times are nearly as good as always selecting the best worker.
Thus, our algorithm should be used instead of the other presented approaches.

\subsection{CPU, Memory, and Bandwidth Utilization}

\begin{figure}[t]
    \centering
    \includegraphics[width=.90\columnwidth]{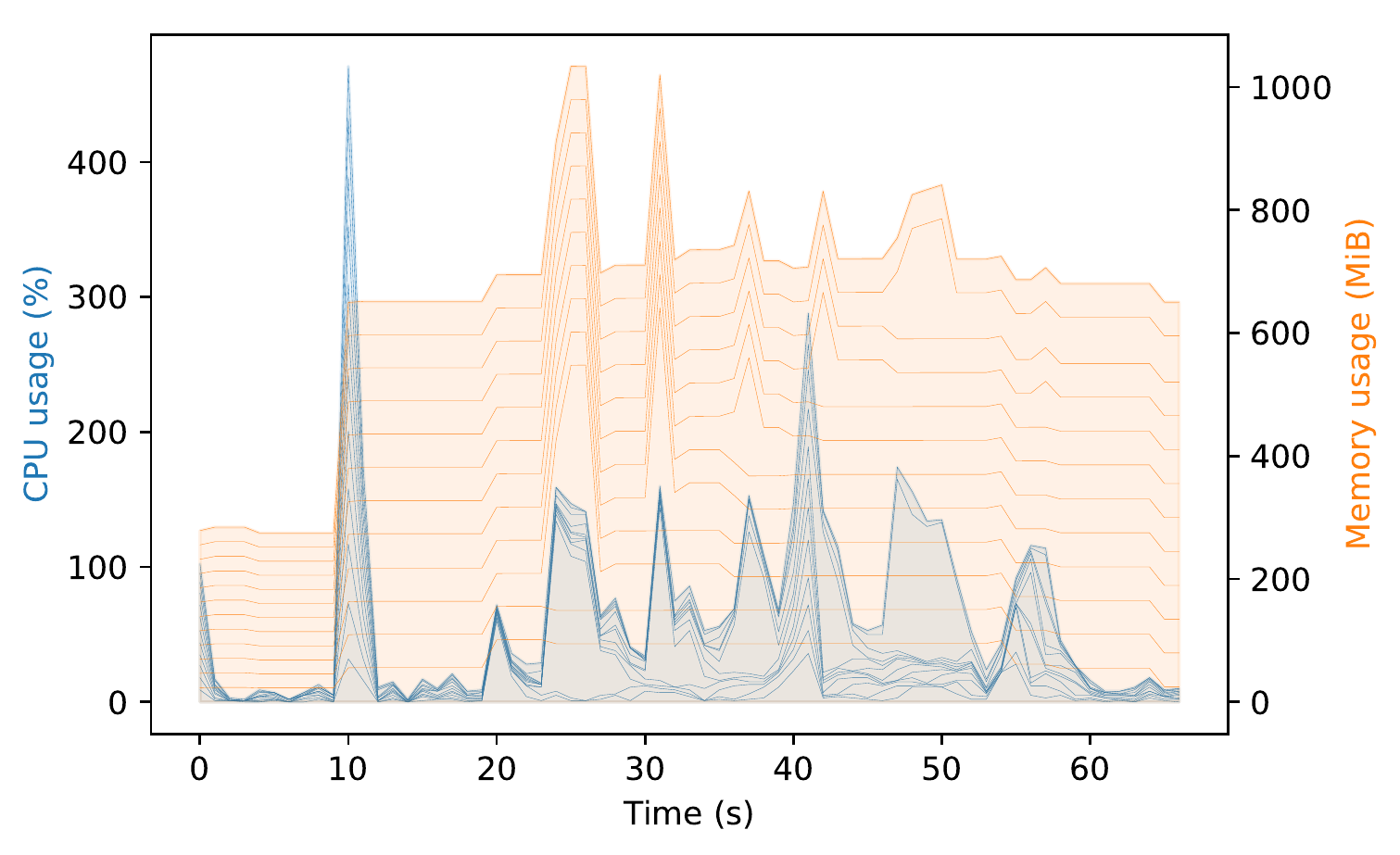}
    \caption{CPU and memory utilization in AoT mode; every worker capable.}
    \label{fig:load}
\end{figure}

Fig. \ref{fig:load} shows the CPU and memory utilization of an experiment in AoT mode where every worker was equally capable to execute a task.
On the x-axis, the time is shown, whereas the left (blue) y-axis denotes the CPU usage and the right (orange) y-axis shows the memory allocation.
In both graphs, the resource usages of all nodes are stacked, whereas 100\% CPU load means that one CPU core of the emulation host is fully utilized (the emulation host had 80 CPU cores and 256 GB RAM, both are not exceeded).

During the first 10 seconds, the test is set up (the emulated nodes are started, configuration files are prepared, etc.).
After 10 seconds, Serval and \dtn are started, which require many computations (e.g., loading Python interpreters into memory, computing hashes for the worker capabilities), and the CPU utilization has a high peak with more than 400\% CPU.
During the experiment, five peaks can be identified, which are the five tasks of the workflow.
The CPU peaks are more blurred, since not only during the task the CPU is used heavily, but also during transmitting the result to the next worker using Serval.
Memory usage shows that on average every node requires about 60 MB of memory, while the execution of a task leads to peaks, due to the fact that the image and the task binary itself have to be loaded into memory.

\subsection{\dtn in Action}
In the final set of experiments, we studied a 30 node network using five different random-waypoint mobility models, since randomly moving nodes is the most challenging scenario in opportunistic networks.
The worker capabilities were set differently in all experiments, as defined in Section~\ref{sub:baseline-eval}.
Furthermore, we evaluated the behavior with 5 and 10 clients that offload tasks at the same time in the network at the start of an experiment, which can lead to workers executing multiple tasks simultaneously.
To simulate an IEEE 802.11g network, which is still widely used especially in decolping countries, with a bandwidth of 54 Mbit/s, a basic range model for the Wi-Fi nodes with 40 meters of range was used.
The mobility model was configured for 30 nodes, walking randomly in an area of about 1.7 km$^2$ at a speed between 0.8 m/s and 1.9 m/s or rest for up to 60 seconds, which corresponds to human walking speed.
This setup leads to relatively small mesh networks that are appearing and disappearing during the execution of the experiment.
Overall, 200 experiments were executed, all using JiT mode.
An experiment finished either successfully, meaning that all clients received their results, or it was stopped after 30 minutes.

\begin{figure}[t]
    \centering
    \includegraphics[width=.90\columnwidth]{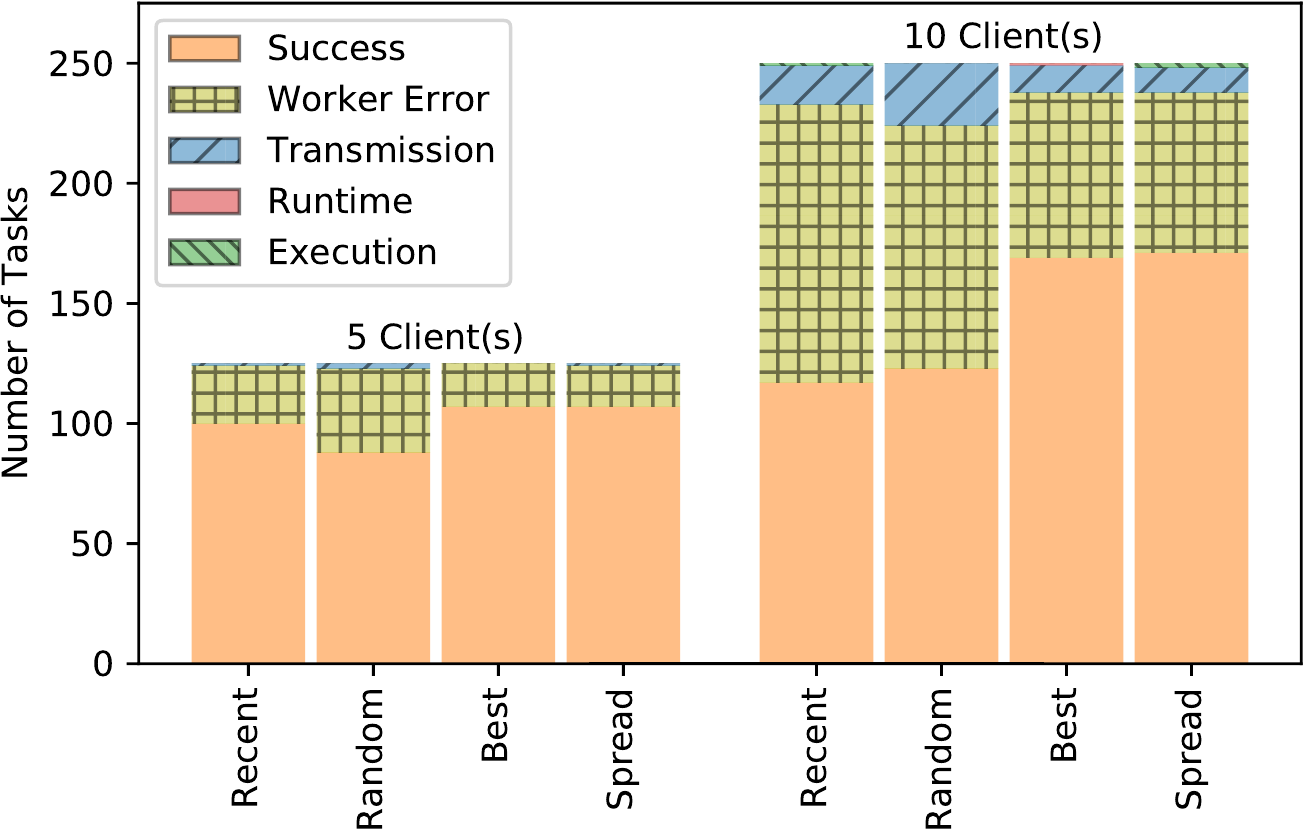}
    \caption{Final workflow states, by number of active clients in JiT mode.}
    \label{fig:mobile-final-states}
\end{figure}

Fig.~\ref{fig:mobile-final-states} shows the final states of the workflows executed in the specific scenarios, where the bars are grouped by the number of clients per experiment and worker assignment.
The y-axis shows the number of tasks in a particular state.
The first case is a successful workflow (\emph{Success}), where a workflow was offloaded, all tasks could be executed, and the result arrived at the client.
Second, \dtn performed as intended but errors occurred as discussed in Section \ref{sub:dealing-with-errors} (\emph{Worker Error}) and the client could successfully be informed about this error.
An experiment stopped in the \emph{Transmission} state, if a task was transmitted to the next worker, but not received until the end of the experiment, e.g., if the recipient cannot be reached due to network fragmentation.
Due to experiment abortion while \dtn was in a runtime state or a worker executed the task itself, it is denoted as \emph{Runtime} and \emph{Execution}, respectively.

Experiments using the recent assignment mode have the lowest success rates, which is due to the fact that workers are selected that are far away and the offers arrive late.
Using a random worker increases the number of successes slightly.
Using the best worker available, all tests were either successful or the client was informed about an error when 5 clients are used.
Our spreading approach is as good as using the best worker in terms of successful workflows or errors returned in time.
The fact that even using the best worker does not lead to 100\% successful executions is due to the worker capabilities and the transmission time in opportunistic networks.
A worker updates its capabilities after executing a task, which can lead to the situation that another task is offloaded to the worker, even though it is not capable anymore.
The falsely assigned worker will decline task execution and inform the client.

\begin{table}[t]
    \centering
    \caption{Average runtimes of tasks in mobile JiT scenarios in seconds.}
    \label{tab:runtimes_mobile}
    \begin{tabularx}{\columnwidth}{Xrrrr}
    \hline
    Assign. & Exec. (s)  & Runt. (s)  & Transm. (s)    & Total (s)      \\
    \hline
    Recent  & 8.7 (0.64) & 5.0 (1.89) & 269.0 (336.37) & 282.8 (338.91) \\
    Random  & 8.9 (1.02) & 5.0 (1.84) & 254.9 (300.75) & 268.8 (303.61) \\
    Best    & 8.9 (0.62) & 5.2 (1.80) & 135.5 (191.26) & 149.6 (193.68) \\
    Spread  & 8.9 (0.68) & 5.1 (1.95) & 234.2 (300.75) & 248.1 (303.61) \\
    \hline
    \end{tabularx}
\end{table}

Table~\ref{tab:runtimes_mobile} shows the average workflow runtimes over all mobile experiments.
It is evident that using our spread algorithm gives better results than random assignment and using a recent worker.
Note that the transmission times (and thus also the total times) have a rather high standard deviation. This is due to the mobility of the nodes and potentially disappearing links between two nodes, resulting in re-transmissions. These increase the time, whereas many transmissions are successful within the first try, reducing the mean transmission time.

To summarize, \dtn introduces negligible overhead in terms of CPU load or memory consumption and supports efficient offloading of computational workflows on resource-constrained devices in opportunistic networking scenarios.

\section{Conclusion}
\label{sec:conclusion}
We presented \dtn, a novel framework for offloading computational workflows in opportunistic networks, with two addressing modes, workers publishing their capabilities and available resources, a worker assignment algorithm, appropriate error handling, and network cleanup to reduce network load.
Experiments with up to 30 emulated mobile nodes showed that worker assignment is important for speeding up workflow execution and for spreading the load fairly on spatially close but powerful workers, which increases the rate of successful offloadings significantly.

There are several areas for future work.
Incorporating further network or social knowledge could improve worker assignment further.
Using \dtn in the field of \emph{Named Data Networking} (NDN) could be interesting for the JiT addressing mode, since clients or intermediate workers would not need to do the worker assignment, but use the abstraction NDN introduces to achieve the same results.
Finally, instead of task chains, support for executing independent tasks of a workflow using DAGs could broaden the scope of \dtn.

\section*{Acknowledgment}
This work has been funded by the German Research Foundation (DFG)
in the Collaborative Research Center (SFB) 1053 MAKI and by the LOEWE initiative (Hessen, Germany) in the NICER and emergenCITY projects.

\bibliographystyle{IEEEtranS}
\bibliography{IEEEabrv,literature}

\end{document}